\documentclass[proceedings]{JHEP3} 

\PrHEP{PrHEP hep2001}%
\usepackage{epsfig}			
\conference{International Europhysics Conference on HEP}
\newbox\mybox%
\newcommand\fverb{\setbox\mybox=\hbox\bgroup\verb}%
\newcommand\fverbdo{\egroup\medskip\noindent\fbox{\unhbox\mybox}\ }%
\newcommand\fverbit{\egroup\item[\fbox{\unhbox\mybox}]}%

\title{Emitter size as a function of mass and transverse mass}%
\author{\speaker{Gideon Alexander} 
\thanks{Talk given at the Int. Europhysics Conf. on HEP, 
July 12-18, 2001, Budapest, Hungary.}\\ 
	Physics Department, Tel-Aviv University, Tel-Aviv, Israel\\%
	E-mail: \email{alex@lep1.tau.ac.il}}%
\abstract{Bose-Einstein and Fermi-Dirac correlations show that
the emitter dimension
$r$ decreases as the hadron mass increases. Same behaviour is 
seen for the longitudinal dimension $r_z$ dependence on the transverse mass
$m_T$. In both cases the Heisenberg
uncertainty relations yield the same expression for $r(m)$ and 
$r_z(m_T)$. This $r$ behaviour also describes the interatomic
separation of Bose condensates. If $r$ represents the emitter 
radius then its energy density reaches for baryon masses 
the high value of $\sim$100 GeV/fm$^3$.  
}%
\begin{document} %
%
%
%
One dimensional (1-D) Bose-Einstein correlations (BEC) of identical 
bosons, and in particular
the pairs $\pi^{\pm}\pi^{\pm}$, have been utilised over several decades
to estimate the emitter size. These analyses used in many cases
the kinematic 
variable $Q=\sqrt{-(q_1 - q_2)^2}$ where $q_i$ are the four momenta
of the two identical bosons. As $Q \to 0$ a BEC
enhancement can be observed in the experimental distribution by comparing
it to a similar distribution void of BEC like e.g., a Monte Carlo 
generated event sample.
The ratio of these two distributions is then described
by the correlation function $C_2(Q)=1 + \lambda e^{-Q^2r^2}$ to yield a value
for $r$, which is taken to be the emitter dimension. 
The factor $\lambda$, known
as the chaoticity parameter, measures the strength of the BEC effect
and can assume the values $0 \leq \lambda \leq 1$.\\

More recently it has been proposed \cite{lipkin} to extract
a similar emitter dimension
for pairs of equal baryons by utilising the so called Fermi-Dirac
correlations (FDC),  
that allows identical fermions at very near phase space, when they
are in an s-wave, only to be in a total spin S=0 state (the Pauli
exclusion principle). To this end a method has been proposed in
reference \cite{lipkin} for the direct measurement,
as function of $Q$, of the fraction of $[S=1]/([S=0]+[S=1])$ in pairs of 
spin 1/2 weakly 
decaying baryons, like the $\Lambda \Lambda$ system. 
Alternatively one can apply the method used in BEC of identical
bosons 
and look at the distribution of baryon pairs as $Q$ approaches zero.
If a depletion is observed then, by assuming its origin to be due to the 
Pauli exclusion principle, an $r$ value can be deduced. The measured baryon 
$r$ values can directly be compared to those obtained for bosons as 
they also measure the distance between the two 
hadrons as the set on of a pure s-wave
state occurs when they approach threshold.\\

The existing vast data of hadronic $Z^0$ decays, three to four million
per LEP experiment, provide an excellent material for BEC and
FDC studies at the same $\sqrt{s_{ee}}$ and in a  
high multiplicity, $\langle n_{ch}\rangle \simeq$21 hadrons,
final state. In particular it was
possible to measure $r$ as a function of the hadron mass. The results
of these analyses are shown in Fig. 1 
where average LEP $r$ values \cite{alex1} are given
for charged pion and kaon pairs, for $\Lambda$ pairs in addition to an
OPAL preliminary $r$ value \cite{letts} for antiproton pairs.  
\DOUBLEFIGURE[htbp]{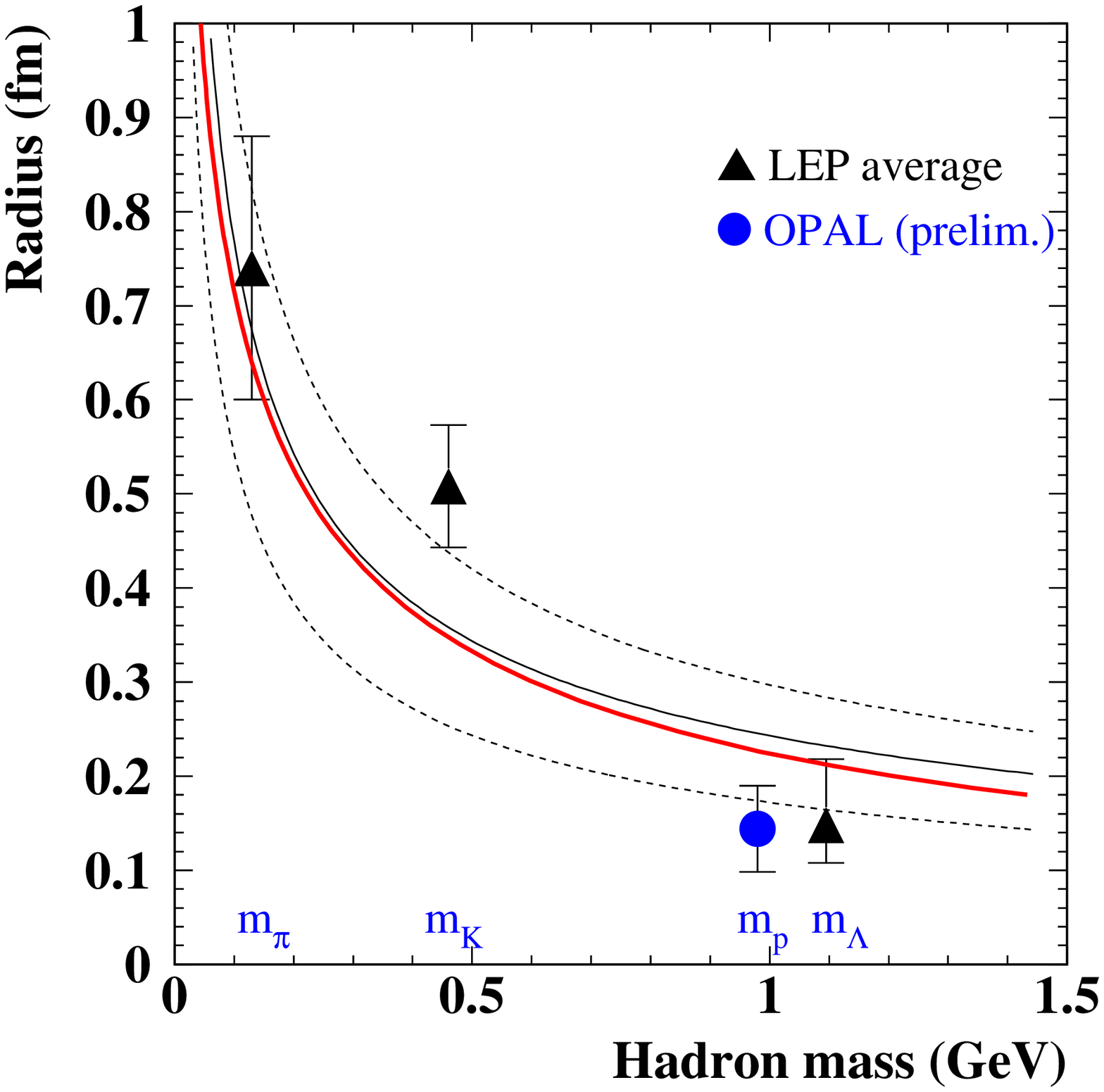,width=8cm}{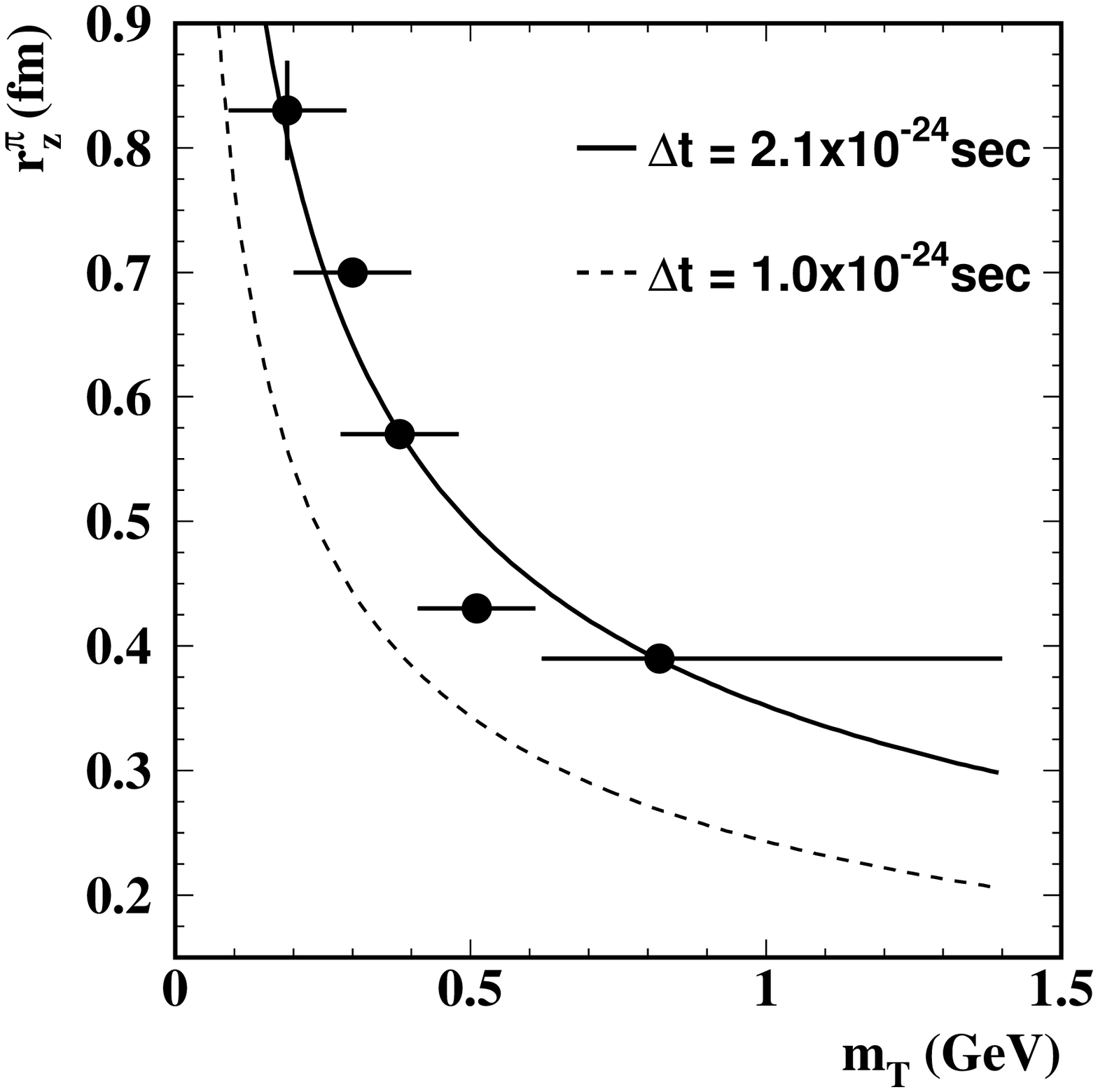,width=8cm}
{The $r(m)$ values (triangles) obtained from 1-D BEC analyses of the hadronic
$Z^0$ decays at LEP \cite{alex1} and an OPAL preliminary 
\cite {letts} value (circle)
for antiprotons. The thin lines are from Eq. 1 
for $\Delta t$ values of $10^{-24}$ sec (central thin line) and
0.5x$10^{-24}$ and 1.5x$10^{-24}$ sec (thin dashed lines).
The thick central line is from the virial theorem using a general QCD
potential \cite{alex1}.}
{Preliminary DELPHI results \cite{delphi_pipi}, obtained from a
2-D BEC analysis, for the  
longitudinal emitter length $r^{\pi}_z$ dependence on the transverse mass
$m_T$ in $Z^0$ decays. The solid and dashed lines are from
Eq. 1 using $\Delta t$ values of 2.1x$10^{-24}$ and
1.0x$10^{-24}$ sec respectively.}
As seen, the average $r$ values decreases from 
$\sim$0.75 fm for
pions down to $\sim$0.15 fm for antiprotons and $\Lambda$ hyperons. 
Whereas the experimental findings that $r(m_{\pi})$ is somewhat larger than
$r(m_K)$, but equal within errors, may still be consistent
with the string fragmentation model although in its basic form it expects
$r(m)$ to increase with $m$, the much smaller value obtained for
$r(\Lambda)$ and $r(antiproton)$ poses a challenge to the model 
\cite{andersson}. On the other hand it has been shown
\cite{alex1} that by applying the Heisenberg uncertainty relations one
can derive an expression for $r(m)$ that decreases with $m$, namely:
\begin{equation}
r(m)\ =\ \frac{c\sqrt{\hbar\Delta t}}{\sqrt{m}}\ .
\label{eq_m}
\end{equation}
The prediction of Eq. \ref{eq_m} is drawn in Fig. 1 and is seen
to follow the general
trend of the experimental values when $\Delta t$ is set to
$\sim 10^{-24}$ sec, to represent a typical time scale for strong
interactions.\\

The effective range of two-pion source was also estimated in 2-dimensional 
(2-D) BEC analyses, in hadronic interactions as well as in the
hadronic $Z^0$ decays \cite{delphi_pipi,l3_pipi}, as a function of the
pion-pair transverse mass $m_T$. This transverse mass is defined as\ 
$m_T=0.5\times\sum_{i=1}^2 \sqrt{m^2+p^2_{i,T}}$ where $p^2_{1,T}$ and
$p^2_{2,T}$ are the transverse momenta of the two bosons defined in the 
longitudinal centre of mass system (LCMS) \cite{csorgo}. The longitudinal
and transverse dimensions $r_z$ and $r_T$ are then obtained from a fit 
of an expression of the type 
$C_2(Q_z,Q_T)=1+\lambda e^{-r^2_zQ^2_z+r^2_TQ^2_T}$ to the data. 
The DELPHI preliminary results \cite{delphi_pipi} for the longitudinal
dimension $r_z$ of two identical charged pion pairs
are seen in Fig. 2 to depend on $m_T$
in a very similar way to the $r(m)$ dependence on 
$m$ (see Fig. 1). In fact, when substituting in Eq. 1, $r$ by $r_z$ and $m$
by $m_T$ one obtains the lines shown in Fig. 2 for two chosen values of
$\Delta t$. This similarity can be understood if one remembers 
that $r_z$ and the 
longitudinal momentum $p_z$ are conjugate observable \cite{alex2}. 
Thus one has 
$\Delta p_z\Delta r_z=2\mu v_zr_z=p_zr_z=\hbar c$ 
where $\mu$ is the reduced mass
of the two hadrons, so that 
\begin{equation}
r_z=\hbar c/p_z\ .
\label{pr_trans}
\end{equation}
Simultaneously we can also use the uncertainty relation given in energy and 
time i.e., $\Delta E \Delta t =\hbar$, where the energy is given in GeV
and $t$ in seconds utilising the fact that in the LCMS,\ 
$p_{1,z}=-p_{2,z}$. 
In as much that the total energy of the boson-pair system
is predominantly determined by the sum of their relativistic mass
values, one has  
$$E = \sum_{i=1}^2 \sqrt{m^2 + p^2_{i,x} +  p^2_{i,y} +  p^2_{i,z}} =
\sum_{i=1}^2 m_{i,T} \sqrt{1 +\frac{p_z^2}{m^2_{i,T}}}
\approx \sum_{i=1}^2\left (m_{i,T} + \frac{p_z^2}{2 m_{i,T}}\right )\ ,$$
where $m_{1,T}$ and $m_{2,T}$ are the transverse mass of the first and
second hadron. At small $Q_z$, the difference 
$\delta m_T=|m_{1,T}-m_{2,T}|/2$ is much smaller than
the transverse mass $m_T=(m_{1,T}+m_{2,T})/2$, and therefore can be 
neglected, so after a few algebraic steps one obtains 
$E=2m_T+p^2_z/m_T$. Since $2m_T$ is not a function of $Q_z$ it may be 
considered to stay fixed as $Q_z \to 0$ so that
\begin{equation}
\Delta E \Delta t \approx (p^2_z/m_T)\times \Delta t = \hbar\ .
\label{et_trans}
\end{equation}
Combining Eqs. \ref{pr_trans} and \ref{et_trans} one finds
\begin{equation}
r_z(m_T)\ \approx\ \frac{c\sqrt{\hbar \Delta t}}{\sqrt{m_T}}\ ,
\label{finalmt}
\end{equation}
which is identical to Eq. \ref{eq_m} when replacing $r$ and $m$ 
respectively by $r_z$ and $m_T$. Here it is worthwhile to note
that in heavy ion collisions it was found out \cite{heinz} that 
$r_z \approx 2/\sqrt{m_T(GeV)}$.  
Experimentally a decrease of
$r_T$ with the increase of $m_T$ was also observed \cite{l3_pipi}
but unlike $r_z$ which is a geometrical quantity,
$r_T$ is a mixture of the transverse radius and the emission time
so that an application of the uncertainty relations is not
straightforward.
An alternative approach for the 
description of $r_z(m_T)$ can be achieved by the so called 
Bjorken-Gottfried conjecture that the momentum-energy 4-vector, $q_{\mu}$, 
is proportional to the space-time 4-vector, $x_{\mu}$. 
In this method one did find  \cite{bialas} that 
$r_z(m_T)$ moves from a typical value of
$\sim 1.1$ fm for $m_T=0.14$ GeV to $\sim 0.25$ fm 
for an $m_T$ of about 1 GeV.\\

Another consequence of the Bose-Einstein statistics of identical
bosons is the existence of Bose condensates of bosonic atoms. These  
condensates, which have been discovered in 1995, are formed by bosonic
atoms when cooled down to temperatures in the typical range of 
500nK to 2$\mu$K, bellow a critical temperature $T_B$,
where the interatomic separation, $d_{BE}$, is of the order of the
de Broglie wave length, $\lambda=\sqrt{h^2/(2\pi mkT)}$. 
Specific calculations \cite{alex2} show that at a very low temperature
$T_0$ where $T_0/T_B \ll 1 $, the average $d_{BE}$ is equal to
\begin{equation}
d_{BE}(m)\ \approx 
\ \frac{\sqrt{2\pi}}{1.378}\left(\frac{\hbar^2}{mkT_0}\right)^{1/2}\ .
\label{dbfinal}
\end{equation} 
From this follows that when two different condensates having atomic mass
$m_1$ and $m_2$ are at the same temperature $T_0$, way below their
individual $T_B$ values, 
the ratio of their interatomic separation will be equal 
to $d_{BE}(m_1)/d_{BE}(m_2)=\sqrt{m_2/m_1}$ similarly to the 
dependence of $r$ ($r_z$) on $m$ ($m_T$). It is 
further interesting to note that in as much that it is permissible
to replace,
at very low temperatures, $kT_0$ by $\Delta E$ and use the uncertainty
relation $\Delta E = \hbar/\Delta t$, one derives 
for $d_{BE}(m)$ the expression given in Eq. \ref{eq_m} for $r(m)$
multiplied by the factor $\sqrt{2\pi}/1.378$. This similarity between
interatomic separation and emitter dimension may well be traced back to the
close connection between the de Broglie wave length and the 
$\Delta p \Delta x \approx \hbar$ Heisenberg uncertainty relation.
Caution should however be exercised when
trying to relate the Bose condensates to the production of hadrons
at high energy reactions. Common to both systems is their bosonic nature
which allows all hadrons (atoms) to occupy the same lowest energy state. 
Furthermore
the condensates are taken to be in a thermal equilibrium state. 
Among the various models proposed for the hadron production some attempts
\cite{thermal} 
have also been made to explore the application of a statistical thermal-like
models however if these will survive is presently questionable. Finally
condensates are taken to be in a coherent state whereas hadron pairs
systems for which an $r$ value can be measured must be at least
partially not coherent, i.e. $\lambda \neq 0$.\\

In as much that the $r$ values obtained from the 1-D BEC analyses
represent the emitter radius one can further try and
estimate the experimental measured 
energy density, ${\LARGE\mbox{$\epsilon$}_{exp}}$,
of the emitter by dividing the sum of the hadron-pair masses 
by a sphere volume of radius $r$,
that is 
\begin{equation}
{\LARGE\mbox{$\epsilon$}_{exp}}=\frac{2m}{(4/3)\pi r^3}\ .  
\label{dens_ex}
\end{equation}
In Fig. 3 the measured energy density of the emitter of pions, kaons
and baryons
are shown in units of GeV/fm$^3$. The data points are compared 
in the figure by the dashed curves with the values expected when $r$ 
given by Eq. \ref{eq_m} is inserted in Eq. \ref{dens_ex} to give
\begin{equation}
{\LARGE\mbox{$\epsilon$}_{model}}=\frac{3}{2\pi}\frac{m^{5/2}}
{c^3(\hbar\Delta t)^{3/2}}\ .  
\label{dens_mo}
\end{equation}
\newpage
\EPSFIGURE[ht]{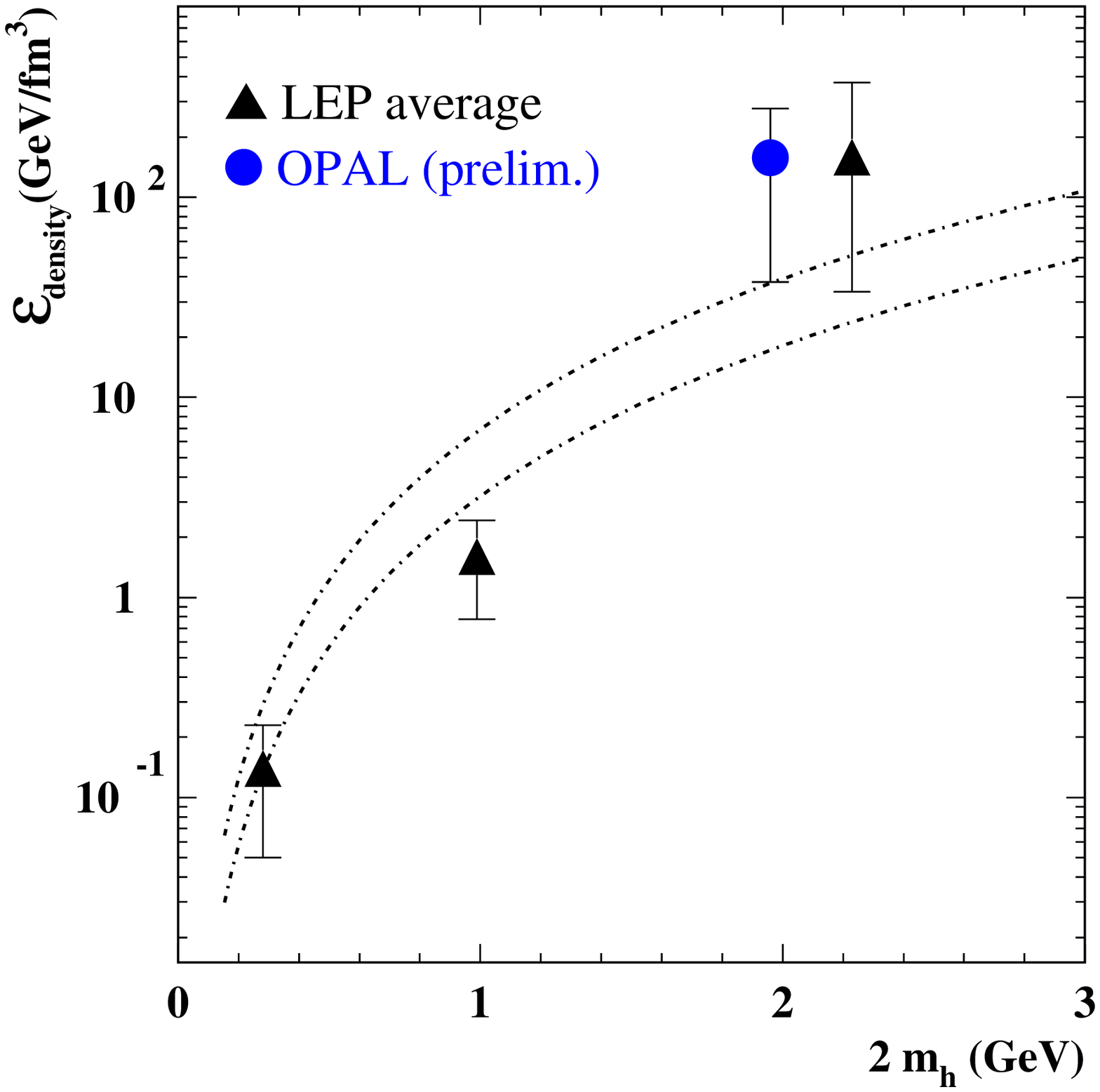,height=6cm}{Emitter energy density
as a function of the hadron-pair mass sum.
The dashed lines are the expectation of 
Eq. \ref{dens_mo} with $\Delta t$=(1.2 $\pm$ 0.3)x$10^{-24}$ sec.}

As can be seen, the energy density values for kaon and pion pairs
are lying in a reasonable range of $\sim$1 GeV/fm$^3$ and below.
On the other hand the energy density of the baryon pairs reaches
an average value of the order of 100 GeV/fm$^3$, very high even in comparison
to the energy density required for the formation of a quark-gluon plasma.
A similar energy density evaluation
of the hadron emitter, deduced from 
2-D BEC analyses, is problematic if not only for the fact
that $r_T$ is
not a pure geometrical quantity.  
In as much that $r$ does in fact represent the
emitter radius then the resulting high energy density poses a 
challenge to the
existing production and hadronisation models 
for hadrons and in particular baryons, emerging from high energy collisions.
  

\acknowledgments{
I would like to thank 
I. Cohen for her help in 
preparing this conference talk and its written version. My thanks are
also due to the 
DESY/Zeuthen laboratory  and its staff for the kind hospitality
extended to me while completing this work.}


\begin{thebibliography}{99}
%
\bibitem{lipkin} G. Alexander, H.J. Lipkin, \plb{352}{1995}{162}. 
\bibitem{alex1} G. Alexander, I. Cohen, E. Levin, \plb{452}{1999}{159}.
\bibitem{letts} OPAL Collaboration, {\it Fermi-Dirac Correlations Between 
Antiprotons in Hadronic Z$^0$ Decays}, PN-486 (2000),  
submitted to this conference, \# 173. 
\bibitem{andersson} B. Andersson, {\it Some remarks on Bose-Einstein 
correlations}, Proc. XXXV Recontres de Moriond, 
QCD, March 18-25, 2000, Les Arcs, France, in press.
\bibitem{delphi_pipi} B. L\"{o}rstad, O.G. Smirnova, {\it Transverse mass
dependence of Bose-Einstein correlation radii in $e^+e^-$ annihilation at
LEP energies}, Proc. 7th Workshop on Multiparticle Production: Correlations
and Fluctuations, Nijmegen 1996, World Scientific, 1997, p. 42.
\bibitem{l3_pipi} L3 Collaboration, {\it A Two-Dimensional Study of
Bose-Einstein correlation in Z decays at LEP}, Nijmegen report HEN-408,
presented at the XXIX Int. Conf. on HEP, Vancouver, 1998.
\bibitem{csorgo} T. Cs\"{o}rg\H{o}, S. Pratt, Proc. Workshop on
Relativistic Heavy-Ion Physics, eds.  Cs\"{o}rg\H{o} et al.,
(KFKI-1991-28/A. Budapest, 1991) p. 75.
\bibitem{alex2} G. Alexander, \plb{506}{2001}{45} and references therein.
\bibitem{heinz} See e.g., U. Heinz, B.V. Jacak, {\it Ann. Rev.
Nucl. Part. Sci.} {\bf 49} (1999) 529 and references therein.
\bibitem{bialas} A. Bialas et al., \prd{62}{2000}{114007} and references
therein. 
\bibitem{thermal} See e.g., F. Becattini et al., 
{\it Nucl. Phys. Proc. Suppl.} {\bf 92} (2001) 137 and references therein.
\end{thebibliography}
\end{document}